\documentclass[]{revtex4-2}

\usepackage{graphicx}
\usepackage{dcolumn}
\usepackage{bm}

\begin{document}

\preprint{AIP/123-QED}

\title[Measuring the mechanical properties of asymmetric membranes ]{Measuring the mechanical properties of asymmetric membranes in computer simulations - new methods and insights}

\author{Oded Farago}
 \altaffiliation{Biomedical Engineering Department, Ben Gurion University of the Negev, 84105, Beer Sheva, Israel}
 \email{ofarago@bgu.ac.il}
%


\date{\today}

\begin{abstract}
We present Monte Carlo simulations of an ultra coarse-grained lipid bilayer with different number of lipids on both leaflets. In the simulations, we employ a new method for measuring the elastic parameters of the membrane, including the area per lipid, area elasticity modulus, and bending rigidity. The method also allows to measure the spontaneous curvature and non-local bending modulus, which are not accessible by standard computer simulations with periodic boundary conditions. For membranes with lipid densities much smaller than the liquid to gel transition density, $\rho_g$, we find a very good agreement between the simulation results and the theory expressing the bilayer elastic free energy as the sum of quadratic free energies in the strains associated with the area density and the local curvature of the monolayers. The theory fails when the lipid area density (in the symmetric reference case) is only slightly smaller than $\rho_g$. Increasing the degree of asymmetry and changing the density of the condensed leaflet to a value larger than  $\rho_g$, causes the layer to phase separate between regions with distinct densities which, in turn, may also induce density variations in the dilated liquid layer. Moreover, the phase separation may also trigger local curvature variations along the membrane, which can be attributed to the disparity between the values of the elastic parameters of the coexisting bilayer segments that are mechanically coupled. This mechanism leading to density-curvature variations and instabilities may play a role in cellular processes occurring in liquid-ordered raft domains that are surrounded by the disordered liquid matrix of the cell.
\end{abstract}

\maketitle

\section{Introduction}
\label{eq:intro}

A key feature of biological membranes is their asymmetry~\cite{PABST2024333,10.1042/ETLS20220088}. The plasma membrane (PM) of eukartotic cells, for example, exhibits an uneven distribution of most lipid species between the extracellular and cytoplasmic leaflets, with nearly all of their sphingomyelin and phosphatidylcholine in the former, and the phosphatidylethanolamine, phosphatidylserine, and phosphatidylinositol in the latter~\cite{VERKLEIJ1973178,10.3389/fcell.2016.00155}. Membrane asymmetry is critical to many biological processes like cell signaling~\cite{https://doi.org/10.1038/sj.emboj.7600798}, processes involving budding (endo- and exocytosis)~\cite{ijms21207594}, and apoptosis~\cite{HANSHAW20055035}. While this feature of the PM has been recognized already more than 50 years ago~\cite{bretschter73}, investigations of the biophysical consequences of the asymmetric nature of biological membranes have only recently gathered momentum. This is largely due to modern tools providing detailed information about the composition and properties (e.g., packing and saturation) of lipids in the two monolayers of the PM~\cite{ilevental20}. Simultaneously, we witness the development of new protocols for the preparation of synthetic vesicles with controllable asymmetry~\cite{membranes13030267}, which paves the way for a new set of research investigations addressing the relationship  between composition, physical properties, and biological functions~\cite{london19,doktorova20,DESERNO2024102832}.  

A bilayer membrane with different lipid composition in the two monolayers is clearly out of thermal equilibrium. Maintaining the asymmetric state requires constant energy consumption. In the PM, this is done through ATP-driven action of membrane proteins like  flippases , floppases, and scramblases~\cite{WOS:000276717600019}, which maintain and control the compositional asymmetry by translocating lipids between the layers, thus countering the passive thermodynamic forces driving the system toward equilibrium and equal compositions. The rate of passive lipid transbilayer movement (flip-flop) is slow (hours to days), and this is largely attributed to the free energy barrier involved in moving the hydrophilic headgroups of the lipids through the hydrophobic bilayer core~\cite{POMORSKI201669}. The slow rates of spontaneous flip-flops imply that active translocation by proteins is essential not only for keeping the non-equilibrium asymmetry of the PM, but also to enable remodeling processes that require faster changes in the compositional asymmetry~\cite{SHUKLA2021183534}. In addition to their direct impact on the lipids, proteins contribute to bilayer asymmetry through structural and functional adaptations. The asymmetric shapes and orientations of transmembrane proteins, established during their directional insertion in the endoplasmic reticulum, not only create uneven functional domains across the bilayer but also stabilize lipid asymmetry by interacting with specific lipid species 

In order to understand the mechanical behavior of biological membranes, one clearly must take their compositional asymmetry into account. Generally speaking, the fact that the outer monolayer is packed with saturated lipids makes it more rigid than the inner leaflet, which is largely populated by unsaturated lipids and is more fluid~\cite{WOS:000220358100004,Holthuis2003LipidML}. This mechanical disparity between the two monolayers modifies the flexibility and stability of asymmetric membranes compared to their symmetric counterparts~\cite{10.1042/ETLS20220084,D2SM00618A}, and is critical to many of the above mentioned biological processes (e.g. endo- and exocytosis). One "positive" consequence of the slow rates of spontaneous lipid exchange between leaflets, is the fact there is a sufficiently long time frame for investigating the mechanical behavior of asymmetric biomimetic membranes. Explicitly, if a model membrane does not contain cholesterol (which is rapidly flip-flopping), the initial asymmetric distribution of the lipids is preserved long enough to enable measurements of quantities of interest characterizing  their elastic response. This is the case, for example, in atomistic simulations or coarse-grained (CG) models like MARTINI~\cite{WOS:000560522700001,membranes13070629}, but not in ultra CG simulations of models like the three-bead implicit-solvent Cooke-Deserno (CD) model~\cite{10.1063/1.2135785}. Such ultra CG models provide a correct quantitative description of the elastic response of lipid bilayers (bending and stretching moduli), but have unphysical large flip-flop rates. This poses no problem in simulations of symmetric membranes, but prevents simulations of asymmetric ones. To overcome this problem, the CD model was amended recently by introducing an additional hydrophobic bead, i.e., making it a four-bead model, which reduces the flip-flopping rates exponentially~\cite{WOS:000592392800036}. 

Here, we consider the original three-bead CD model and demonstrate how it can be used for simulations of asymmetric bilayers. The key difference between our work and other simulation studies of asymmetric membranes is the fact that we run Monte Carlo (MC) rather than Molecular Dynamics simulations, which allows us to introduce rejection rules that prohibit flip-flopping. The type of asymmetry considered herein is the simplest one: We simulate a membrane consisting of a single lipid type but with different numbers of lipids in the two monolayers~\cite{WOS:000349614200005,HOSSEIN2020624}. With the MC simulations, we introduce a new method for measuring the bending modulus of the simulated bilayers, and validate it by simulating symmetric membranes and comparing our results to published data which is based on other approaches, e.g., Fourier analysis of thermal fluctuations. A novel feature of the method is that although the bilayer is subjected to conventional periodic boundary conditions (PBCs) and is flat on average, the method also allows measurements of the so called non-local bending modulus which is associated with the area-difference elasticity of closed vesicles~\cite{TAKIUE2022101646} (see next section), and which has hardly been measured computationally. In the following section, we briefly review the elastic theory of asymmetric bilayers. The theory is based on the assumption that the two monolayers are fluid and connected in parallel without modifying each other's individual elastic response. Our results are in agreement with this theoretical framework, but only if the model parameters are set such that the {\em symmetric}\/ bilayer is far from the gel transition. Near the gel transition, we observe that the compressed monolayer may undergo a first order phase transition and show coexisting gel and liquid regions with distinct non-vanishing curvatures of opposite sign. This type of curvature instability is expected in raft-forming biological membranes~\cite{schick18}, and is likely to be relevant in several biological processes.     

\section{Elasticity of asymmetric membranes}
\label{sec:theory}

The elastic theory of asymmetric membranes has been discussed in several recent publications~\cite{WOS:000349614200005,HOSSEIN2020624,svetina96} and is briefly reviewed here for completeness. 
We begin by writing the elastic free energy of the bilayer as the sum of the elastic free energies of its constituent monolayers ($i=1,2$): 
\begin{equation}
F=\sum_{i=1}^2\left[\frac{K_i(A_i-A_{i,0})^2}{2A_{i,0}}+\int\frac{\kappa_i}{2}(C_i-C_{i,0})^2dA_i\right].
\label{eq:bilayelasticity1}
\end{equation}
The first term in Eq.~(\ref{eq:bilayelasticity1}) represents the area elasticity of the mono-
layers with $K_i$, $A_i$ and $A_{i,0}$ denoting the stretching/compression
(two-dimensional) modulus, area, and relaxed area of the monolayers, respectively. The second term represents the bending energy, with $\kappa_i$, $C_i$ and $C_{i,0}$ denoting, respectively, the monolayers bending modulus, sum of principle curvatures (twice the mean
curvature - to be henceforth called "the curvature"), and the spontaneous curvature. The Gaussian curvature term has been omitted by virtue of the Gauss-Bonnet theorem. Summing the monolayers contributions in Eq.~(\ref{eq:bilayelasticity1}) and expressing the energy in terms
of the so called neutral surface~\cite{kozlov93}, we can write
\begin{equation}
    F=\frac{K(A-A_0)^2}{2A_0}+\frac{K_r(\Delta A-\Delta A_0)^2}{2A_0}+\int\frac{\kappa}{2}(C-C_0)^2dA.
 \label{eq:bilayelasticity2}   
\end{equation}
The first and the last terms on the r.h.s~of Eq.~(\ref{eq:bilayelasticity2}) represent the area and bending elasticities of the bilayers, and have similar forms to the corresponding monolayers contributions in Eq.~(\ref{eq:bilayelasticity1}) with~\cite{svetina96}
\begin{eqnarray}
     A_0&=&\frac{K_1+K_2}{K_1/A_{1,0}+K_2/A_{2,0}}  \label{eq:a0} \\
     K&=&K_1+K_2 \label{eq:k} \\
     C_0&=& \frac{\kappa_1C_{1,0}+\kappa_2C_{2,0}} {\kappa_1+\kappa_2} \label{eq:c0} \\
     \kappa&=&\kappa_1+\kappa_2 \label{eq:kappa} 
\end{eqnarray}
In the middle term, $\Delta A=A_1-A_2$ is the area difference between the monolayers, and $\Delta A_0=A_{1,0}-A_{2,0}$ is the optimal area difference. This contribution to the energy of closed membranes has been termed {\em area difference elasticity}\/~\cite{seifert97,BIAN2020112758}. Using the parallel surface theorem~\cite{DESERNO201511}, we can relate the area difference to the average curvature, $\overline{C}$, defined by 
\begin{equation}
    \Delta A=h\int CdA\equiv hA\overline{C},
    \label{eq:avcurvature}
\end{equation}
where $h$ is the distance between the surfaces of the two leaflets. Thus, the area difference elasticity term can be rewritten in a form resembling the bending rigidity term, i.e., 
\begin{equation}
    F=\frac{K(A-A_0)^2}{2A_0}+\frac{\kappa_{\rm nl}}{2}(\overline{C}-\overline{C}_0)^2A+\int\frac{\kappa}{2}(C-C_0)^2dA.
 \label{eq:bilayelasticity3}   
\end{equation}
The modulus $\kappa_{\rm nl}$ is the {\em non-local}\/ bending rigidity of the vesicle~\cite{SVETINA2014189}. In contrast to the local bending rigidity $\kappa$, it is not associated with the molecular shape of the lipids and the corresponding preferred curvature, but is linked to elastic frustration arising from the fact that it may not be possible to relax the areas of the two leaflets simultaneously. From Eq.~(\ref{eq:bilayelasticity2}) we conclude that a closed tensionless bilayer ($A=A_0$) may be in a state of {\em differential stress}\/, i.e., its two leaflets may experience non-vanishing stresses of equal size and opposite signs. This can be understood by considering a surface with constant curvature $C=\overline{C}$ (e.g., a sphere or a cylinder). Differentiating $E$ in Eq.~(\ref{eq:bilayelasticity3}) w.r.t.~$\overline{C}$, we find that the energy is minimized when 
\begin{equation}
    C^*=\frac{\kappa C_0+\kappa_{\rm nl}\overline{C}_0}{\kappa+\kappa_{\rm nl}},
    \label{eq:meancurve}
\end{equation}
i.e., for $C^*\neq \overline{C}_0$, which implies that while the bilayer is tensionless, the monolayers are not.

Since $\kappa_{\rm nl}$ is coupled to $\overline{C}$, it can be measured in experiments involving a change in the average curvature like micropipette aspiration~\cite{WAUGH1992974}, but not in flicker experiments analyzing the thermal fluctuation of a quasi-spherical vesicle with a given radius $R\simeq 2/\overline{C}$. The same applies to computer simulations with PBCs, where the average curvature $\overline{C}=0$ is fixed~\cite{Wang_Hu_Zhang_2013}, and the fluctuation spectrum is governed by $\kappa$ but insensitive to $\kappa_{\rm nl}$. In the following section, we describe a computational method for measurements of both $\kappa$ and  $\kappa_{\rm nl}$ in ultra CG membrane simulations of asymmetric membranes.   

\begin{figure}[t]
\centering
  \includegraphics[height=6cm]{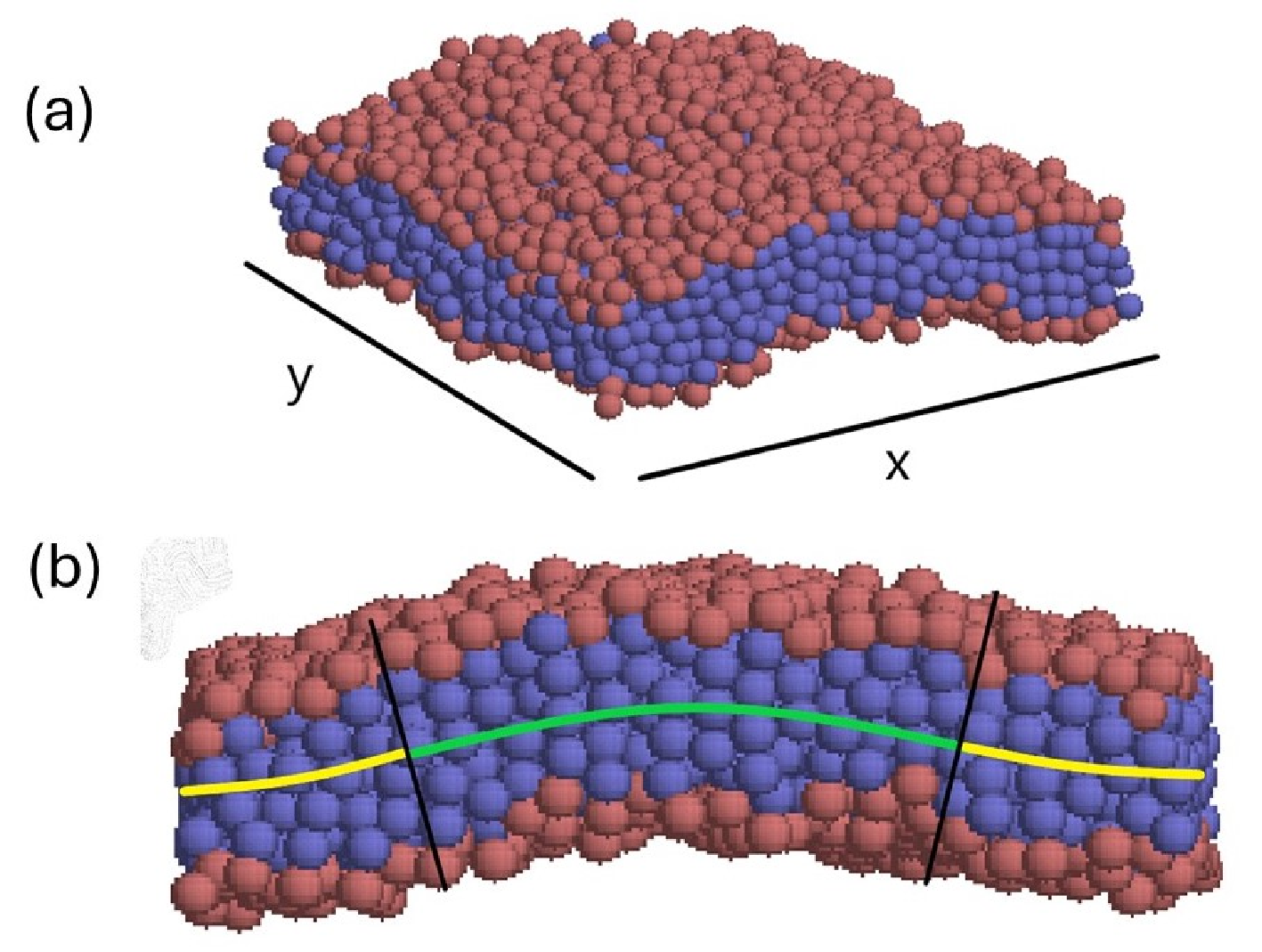}
  \caption{Snapshot of an asymmetric bilayer membrane with different number of lipids in both leaflets. The membrane consists of two cylindrical arcs with opposite curvatures and the same spanning angle, such that they join smoothly to each other. (a) A view from the "corner" with $y$ being the direction parallel to the axes of the cylinders, and $x$ the direction along which the membrane has a "wiggly" shape. (b) A front view with the green and yellow arcs marking the mid-plane of the membrane, and the black lines indicating the interfaces along which the two arcs are joined together.}
  \label{fig1}
\end{figure}  

\section{Computational Methodology}
\label{sec:method}

The method for determining the local and non-local bending rigidities of asymmetric bilayers is based on a useful feature of MC simulations, which is the ability to reject certain moves and thereby to sample a partial configurational phase space. Specifically, we consider membrane configurations similar to the one shown in fig.~\ref{fig1}, consisting of two cylindrical arcs having their axes lying along the $y$ direction - one at $x=0$ and the other in the middle of the simulation box at $x=L_x/2$ [fig.~\ref{fig1}(a)]. The two arcs, which are marked by green and yellow colors in fig.~\ref{fig1}(b), have opposite curvatures $\pm c$ and similar spanning angels $\theta$ and, therefore, they can are joined smoothly to each other along the interfaces indicated by black lines in fig.~\ref{fig1}(b). PBCs are applied along both the $x$ and $y$ directions

The green/yellow arcs in fig.~\ref{fig1}(b) show the mid-plane of the membrane. Above and below the mid-plane are the two leaflets with $N_+$ and $N_-$ lipids, respectively. We use the CD model (see details in ref.~\cite{10.1063/1.2135785}), where each lipid is represented as a trimer consisting of one hydrophilic (red) and two hydrophobic (blue) beads of diameter $\sigma$. In the simulations, we set the CD model parameter $\omega_c=1.35$ and tune the membrane fluidity by varying  the other CD model parameter, $\epsilon$. The position of each lipid $(r_l,\theta_l,y_l)$ is specified by the location of the middle bead, where $r_l$ is the radial distance from the axis of the relevant cylinder, $-\theta/2<\theta_l<+\theta/2$ is the angular coordinate (with $\theta$ being the spanning angle of the arcs), and $y_l$ is the coordinate along the cylinder axis. As noted above, the CD model "suffers" from a very high rate of flip-flops. To maintain the initial asymmetry of the bilayer, we use the following rejection rule to restrict the radial coordinate of the lipids: Consider the snapshot in fig.~\ref{fig1}(b) with positive curvature $c$ in the segment indicated by the green arc, and an opposite negative curvature in the other segment indicated by yellow. The radial coordinates of the lipids in the upper leaflet in the figure must satisfy $1/c<r_l<1/c+2.5\sigma$ in the segment with the positive curvature, and $1/c-2.5\sigma<r_l<1/c$ in the segment with negative curvature. This restriction prevents not only flip-flopping of lipids between the layers, but also their escape out of the membrane and, thus, guarantees the integrity of the membrane over long simulations. 

We simulate {\em tensionless}\/ membranes with $2N_0=N_++N_-=1200$ lipids in a simulation box whose length along the $y$ axis is fixed to $L_y=28.25\sigma$. The simulations involve displacements and rotations of randomly selected lipids, as well as occasional "collective" moves attempting changes in (i) the spanning angle $\theta$, and (ii) the curvature of the two membrane segments $\pm c$ (in which case we change the angle so that $\theta|c|$ remains unchanged, or otherwise the area of the bilayer may change significantly and the  acceptance will be nearly-vanishing). All moves are accepted according to the Metropolis criterion with the addition of the above-mentioned rejection rule for the locations of the lipids. In most of the simulations, the lipids are allowed to move between the two segments with the opposite curvatures. If a lipid is displaced from one segment into the other (and the move is accepted), then it is simply counted as part of the same layer but in the new segment. When both layers are liquid, the lipids diffuse between the two segments and establish a uniform density throughout each layer. In the asymmetric case, the densities will be different between the two layers, i.e., the tensionless bilayer will have opposite stresses in the two monolayers. However, according to Eq.~(\ref{eq:bilayelasticity3}), the differential stress is not expected to modify the mechanical behavior of the simulated membrane because all the sampled configurations have, by construction, $\overline{C}=0$ and, therefore, the non-local bending term is a constant. The importance of the non-local bending term is expected to be negligible also for vesicles with radius $R=1/\overline{C}\gg h$, where $2h$ [see definition of $h$ in Eq.~(\ref{eq:avcurvature})] is the molecular width of the bilayer ($h \sim 2$~nm), as long as they retain their quasi-spherical shape and do not bud. For the simulated phase space, the elastic free energy of configurations with a given area $A$ and curvatures $\pm c$ in the two segments is
\begin{eqnarray}
    F(A,c)&=&\frac{K(A-A_0)^2}{2A_0}+\frac{\kappa}{2}\left[(c-C_0)^2+(-c-C_0)^2\right]\frac{A}{2}\nonumber \\
    &\simeq& \frac{K(A-A_0)^2}{2A_0}+A_0\frac{\kappa}{2}c^2+A_0\frac{\kappa}{2}C_0^2,
 \label{eq:elasticity1}   
\end{eqnarray}
and the last term is obviously an unimportant constant that can be discarded. This result implies that the area and curvature are independent quadratic degrees of freedom since there is no coupling between $A$ and $c$ in Eq.~(\ref{eq:elasticity1}), which reads $F(A,c)=F_1(A)+F_2(c)$ with 
\begin{eqnarray}
    F_1(A)&=&\frac{K(A-A_0)^2}{2A_0}+ E_1, \label{eq:farea} \\
    F_2(c)&=&A_0\frac{\kappa}{2}c^2+E_2. \label{eq:fcurve}
\end{eqnarray}
This means that the elastic moduli of the bilayer can be measured by using the equipartition theorem: $\langle(A-A_0)^2\rangle=A_0(k_BT/K)$, where $A_0=\langle A\rangle$, and $\langle c^2\rangle=(A_0)^{-1}(k_BT/\kappa)$. However, we can do better than that, and extract from the simulations the whole probability distributions of the area and curvature, $\Pi(A)$ and $\Psi(c)$, which if found to be Gaussian would prove the validity of Eqs.~(\ref{eq:farea}) and (\ref{eq:fcurve}). The elastic moduli can then be determined by fitting these equations for the free energies to the logarithms of the probability distributions:
\begin{eqnarray}
    F_1(A)=-k_BT\ln\left[\Pi\left(A\right)\right], \label{eq:fit1}
    \\
    F_2(c)=-k_BT\ln\left[\Psi\left(c\right)\right]. \label{eq:fit2}
\end{eqnarray}

\begin{figure}[t]
\centering
\includegraphics[height=10cm]{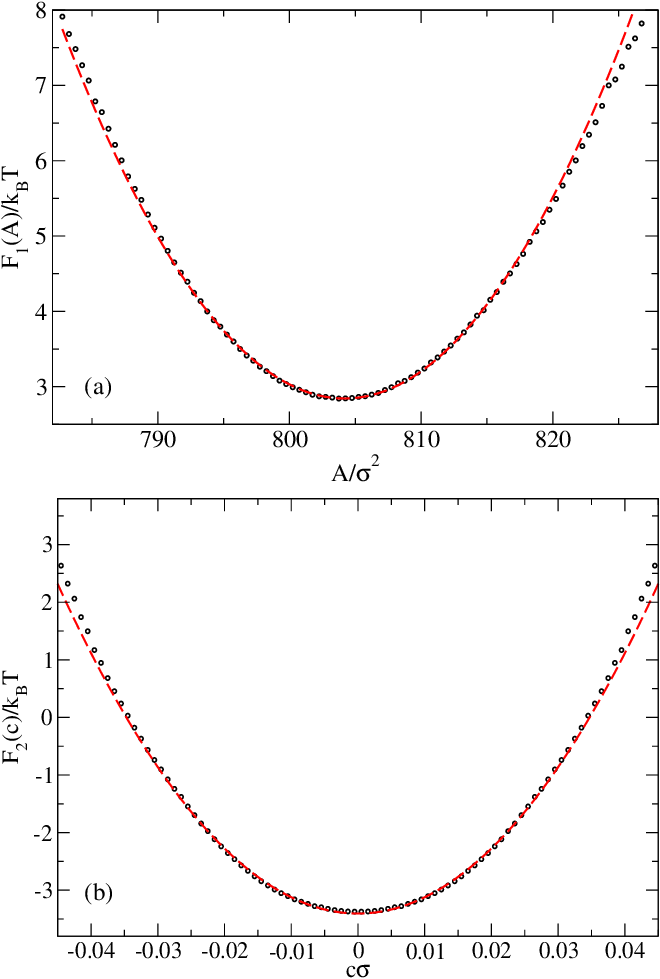}
  \caption{The logarithms of the probability distribution functions of the (a) area $A$, and (b) curvature, of a symmetric membrane. Black circles -  computational results; red dashed lines - fits of the results to the quadratic forms of the free energies Eqs.~(\ref{eq:farea}) and (\ref{eq:fcurve}), respectively.}
  \label{fig2}
\end{figure}

To validate the method, we simulated the symmetric case $N_+=N_-=N_0$ for the CD model parameter $\epsilon=1.05 k_BT$. For this value of $\epsilon$, the bilayer is fluid and very flexible, and in previous simulations of the same system we found $\kappa=8\pm1 k_BT$ based on Fourier analysis of the spectrum of thermal fluctuations~\cite{farago08}. This value of $\kappa$ is characteristic of red blood cells~\cite{rbc22}. Fig.~(\ref{fig2}) shows the MC simulations results for the logarithms of the probability densities of the (a) area - $\Pi(A)$ and (b) curvature - $\Psi(c)$ (black symbols), and their fits to free energies $F_1(A)$ and $F_2(c)$ (red dashed lines). The data fit very nicely to the quadratic forms of Eqs.~(\ref{eq:farea}) and (\ref{eq:fcurve}), with some non-quadratic corrections visible away from the minima of the free energies. These correections at large strains are expected because of the excluded volume interactions causing the response to compression to be stiffer  than the response to stretching. From the fits at small strains, we derive the following values for the equilibrium area and the stretching/compression modulus: $A_0=804~\sigma^2$  (area per lipid $a_0=A_0/N_0=1.340\sigma^2$), $K=(17.2\pm0.2)\,k_BT/\sigma^2$. For the bending rigidity we find $\kappa=(7.0\pm 0.5)\,k_BT$, consistent with the measurement in ref.~\cite{farago08}.

\begin{figure}[t]
\centering
\includegraphics[height=10cm]{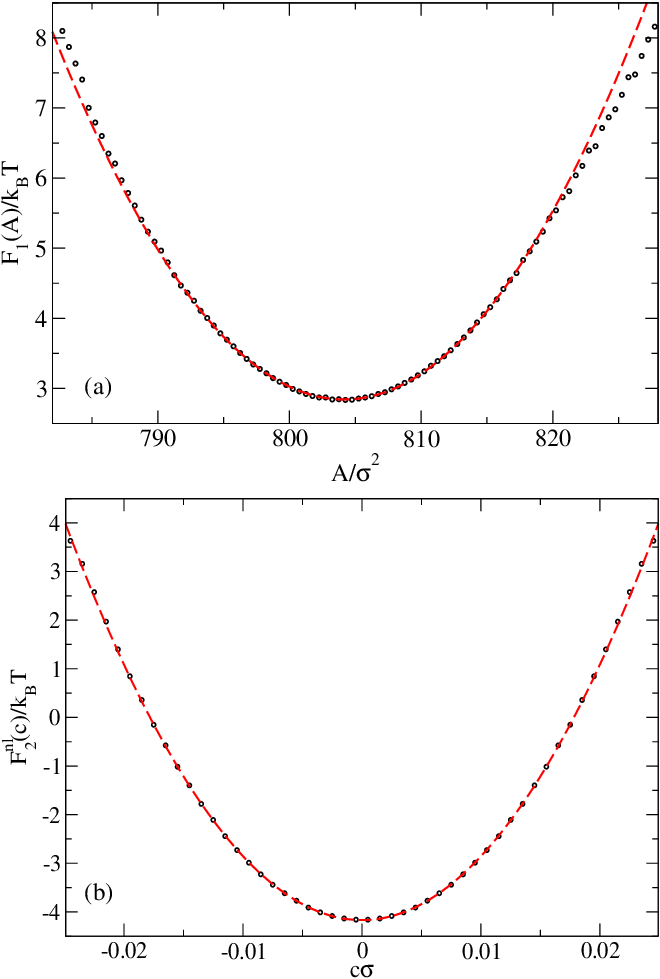}
  \caption{Same as in fig.~\ref{fig2}, but for a symmetric membrane also having non-local bending rigidity. The simulation data (symbols) is fitted to the free energies Eqs.~(\ref{eq:farea}) for the area in (a), and (\ref{eq:fcurvenl}) for the curvature in~(b).}
  \label{fig3}
\end{figure}

\begin{figure}[t]
\centering
\includegraphics[height=12.5cm]{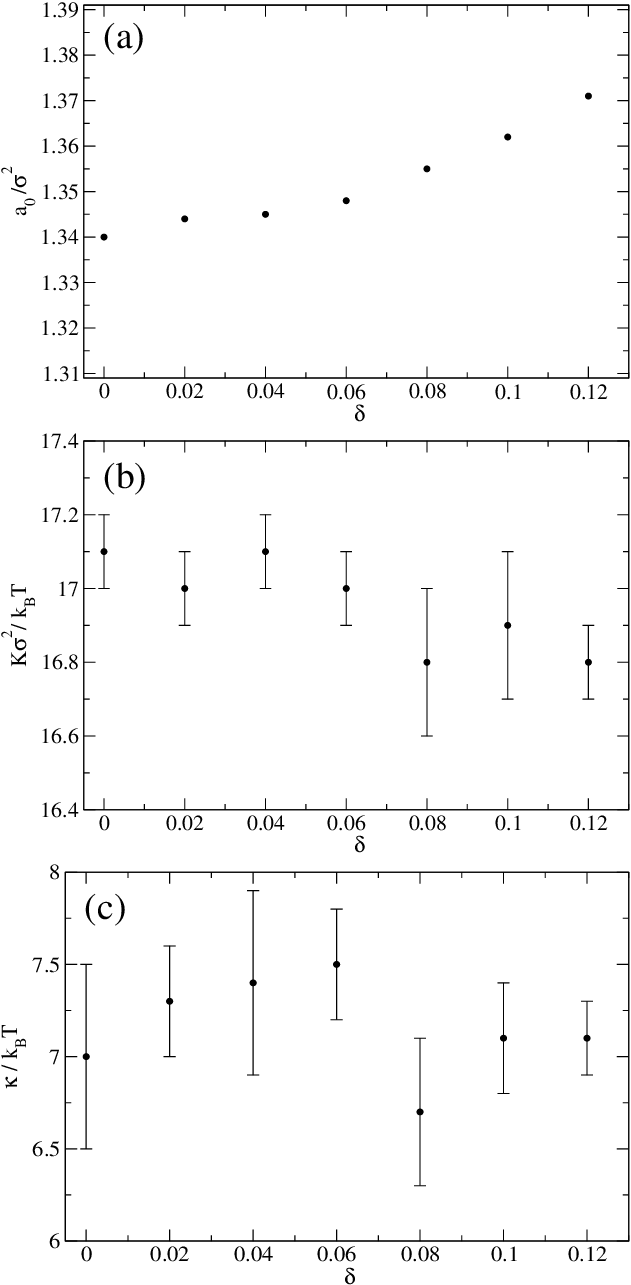}
  \caption{(a) The relaxed area per lipid $a_0=A_0/N_0$, (b) area elasticity modulus $K$, and (c) bending modulus $\kappa$, as a function of the asymmetry parameter $\delta=(N_+-N_-)/2N_0$. The simulated bilayers have CD parameter $\epsilon=1.05\, k_BT$, and the measurements are conducted in the setting without non-local bending rigidity.} 
  \label{fig4}
\end{figure}

We can also measure the non-local bending modulus of the membrane. This is done by introducing an additional rejection rule prohibiting displacement of lipids between the two membrane segments. By initially placing the same number of lipids $N_+$ in the {\em upper}\/ layer of the segment with positive curvature $c$, and in the opposite {\em lower}\/ layer of the segment with the opposite negative curvature [indicated, respectively, by green and yellow arcs in fig.~\ref{fig1}(b)], we create a bilayer composed of two replicas of cylindrical arcs having the same asymmetry. In this configuration phase space, the segments can relax individually into a state where the stresses in both monolayers vanish, i.e., with no differential stress. In the symmetric case ($N_+=N_0$), the bilayer will be on average flat, but in the asymmetric case ($N_+> N_0$), the state of mechanical equilibrium will have a non-vanishing spontaneous curvature. Furthermore, curvature fluctuations around this value will be attenuated by both the local and non-local bending rigidities because they also involve changes in the area difference $\Delta A$ in each of the two segments. 
The free energy in this set of simulations is expected to have the form $F(A,c)=F_1(A)+F_2^{\rm nl}$ (the superscript "nl" indicates that this free energy also involves a non-local contribution) with $F_1(A)$ having the same form as in Eq.~(\ref{eq:farea}), and replacing Eq.~(\ref{eq:fcurve}) with
\begin{equation}
    F_2^{\rm nl}(c)=A_0\frac{(\kappa+\kappa_{\rm nl})}{2}(c-C^*)^2+E_2^{\rm nl} \label{eq:fcurvenl},
\end{equation}
where $c$ is the curvature of the segment with $N_+$ lipids in the upper layer, and  $C^*$ is the spontaneous curvature given by Eq.~(\ref{eq:meancurve}). Fig.~\ref{fig3} displays the analysis of the results of for this set of simulations, similarly to fig.~\ref{fig2}  but with the curvature free energy fitted to Eq.~(\ref{eq:fcurvenl}). We obtain exactly the same values $A_0=804\sigma^2$ and $K=(17.2\pm0.2)\,k_BT/\sigma^2$ for the area parameters, and a total bending rigidity of $\kappa_{\rm tot}=\kappa+\kappa_{\rm nl}=(32.4\pm0.3)\, k_BT$. Since $\kappa$ is already known for this system, we conclude that $\kappa_{\rm nl}=\kappa_{\rm tot}-\kappa\simeq (25.4\pm0.4)\, k_BT$.

\begin{figure}[t]
\centering
\includegraphics[height=8.75cm]{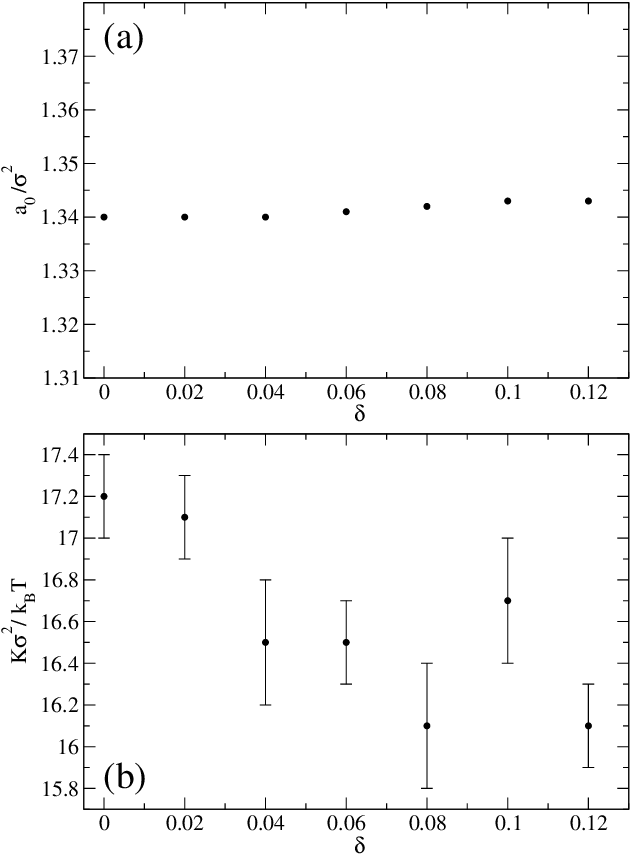}
  \caption{(a), (b) Same as in fig.~\ref{fig4}, but for the simulations setting with non-local bending rigidity.} 
  \label{fig5}
\end{figure}

\begin{figure}[t]
\centering
\includegraphics[height=8.75cm]{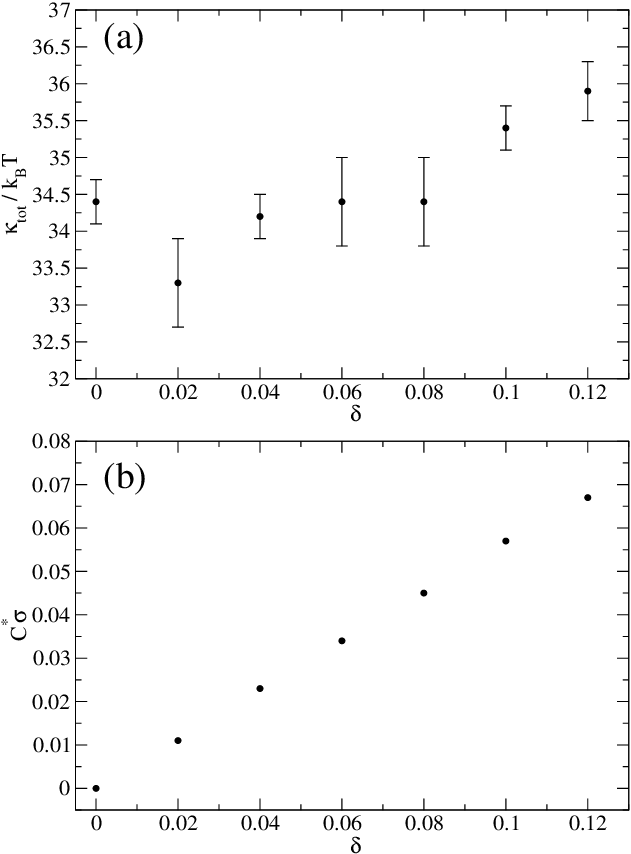}
  \caption{(a) Total bending rigidity, and (b) spontaneous curvature as 
  a function of the asymmetry parameter $\delta$. The results are obtained in the same set of simulations as in fig.~\ref{fig5} (i.e., for the CD parameter $\epsilon=1.05 k_BT$ and in a setting with non-local bending rigidity).} 
  \label{fig6}
\end{figure}

\section{Asymmetric fluid membranes}
\label{sec:fluid}

Fig.~\ref{fig4} shows the variations in the mechanical properties of asymmetric membranes with $N_+$ lipids in the upper layer and $N_-\leq N_+$ in the lower layer, as a function of the asymmetry parameter defined as $\delta=(N_+-N_-)/2N_0$. The simulation results shown here correspond to bilayers with CD parameter $\epsilon=1.05\, k_BT$, which are fluid and flexible, and which were simulated in the setting without non-local bending rigidity (see previous section~\ref{sec:method}). Sub-figures (a), (b), and (c) present the results for the relaxed area per lipid $a_0$, bilayer area elasticity modulus $K$, and bilayer bending modulus $\kappa$, respectively. The agreement with the theoretical framework in section~\ref{sec:theory} is very good. Explicitly, Eqs.~(\ref{eq:a0}), (\ref{eq:k}) and (\ref{eq:kappa}) suggest that if the monolayers properties vary linearly with $\delta$ in opposite manners to each other [i.e., $\sim B_0\pm B_1\delta+\mathcal{O}(\delta^2)$], then the corresponding bilayer properties should have only a weak $\delta ^2$ dependence on the asymmetry parameter. Indeed, fig.~\ref{fig4}(c) shows no significant variations in the bending modulus. The area elasticity modulus and the relaxed area exhibit a slight linear dependence on $\delta$, which can be attributed to the weak non-linear elastic response, also observed at larger areal strains in fig.~\ref{fig2}(a).  

Figs.~\ref{fig5} and \ref{fig6} show the results for the same membrane (with $\epsilon=1.05k_BT$) but in the setting with non-local bending rigidity. The area elasticity parameters, $A_0$ and $K$, are plotted in fig.~\ref{fig5} (a) and (b), respectively. As expected, the relaxed area does not vary with $\delta$, and is identical to the relaxed area derived in fig.~\ref{fig4} for $\delta=0$, i.e., for the membrane with no differential stress. Moreover, according to the theory in section~\ref{sec:theory}, non-local rigidity (area-difference elasticity) represents an elastic deformation mode which is distinct from the area-elasticity and bending rigidity and, therefore, the results for the area elasticity modulus $K$ in fig.~\ref{fig5}(b) should be similar to those displayed in fig.~\ref{fig4}(b) when the non-local rigidity is absent. Comparing the data in these two figures confirms that the measurements for $K$ in both setting are in a fairly good agreement with each other. Fig.~\ref{fig6}(a) shows the total bending rigidity $\kappa_{\rm tot}=\kappa+\kappa_{\rm nl}$. The dependence of this elastic on  $\delta$ is weak, suggesting that both $\kappa$ and $\kappa_{\rm nl}$ vary only slightly with the degree of asymmetry. Fig.~\ref{fig6}(b) shows the dependence of the spontaneous curvature $C^*$ on $\delta$. The displayed linear dependence is fully consistent with Eq.~(\Ref{eq:meancurve}), since $C^*$ is proportional to the optimal area difference between the monolayers: $C^*\propto\Delta A_0\propto\delta$ (provided that $C_0=0$ for $\delta=0$).

\begin{figure}[t]
\centering
\includegraphics[height=10cm]{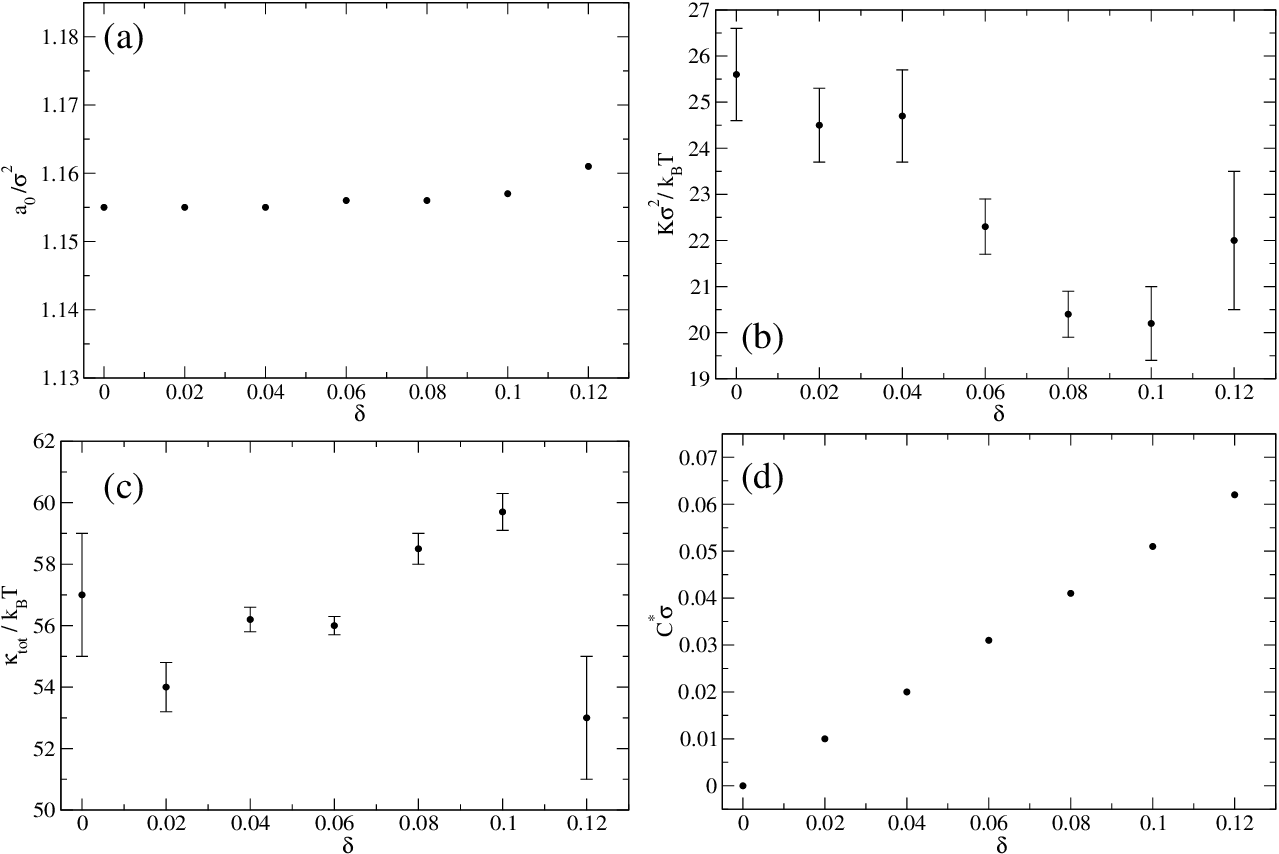}
  \caption{(a) The relaxed area per lipid $a_0=A_0/N_0$, (b) area elasticity modulus $K$, (c) total bending rigidity $\kappa_{\rm tot}$, and (d) spontaneous curvature $C^*$, as a function of the asymmetry parameter $\delta$. The simulated bilayers have CD parameter $\epsilon=1.3\, k_BT$, and the measurements are conducted in the setting with non-local bending rigidity.} 
  \label{fig7}
\end{figure}

\begin{figure}
\centering
\includegraphics[height=5cm]{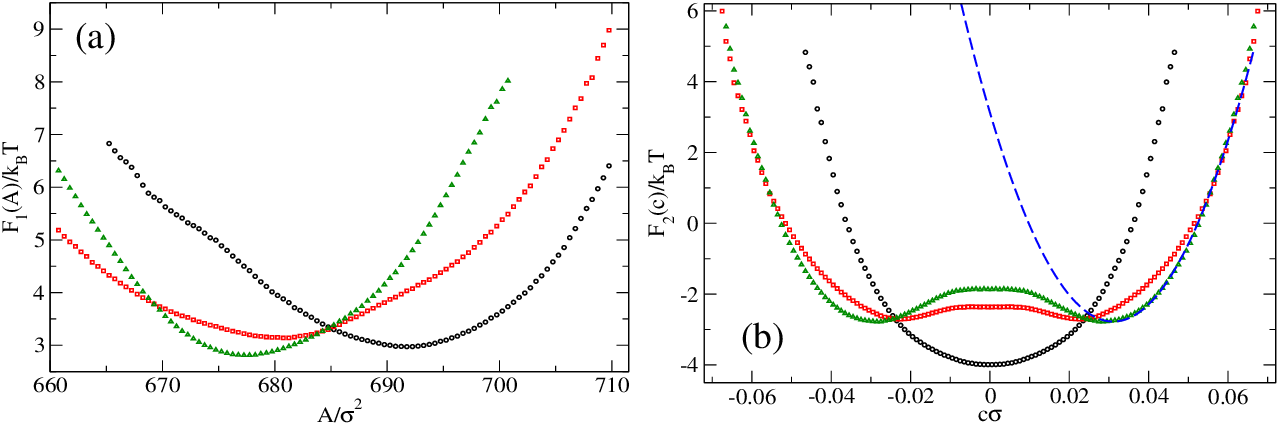}
  \caption{The logarithms of the probability distribution functions of the (a) area $A$, and (b) curvature $c$, for membranes with CD parameter $\epsilon=1.3\,k_BT$ in the setting without non-local rigidity. Black circles, red squares, and green triangles show data for $\delta=0.06,\ 0.08$, and $0.1$, respectively. The blue dashed line in (b) is a fit of the green symbols to Eq.~(\ref{eq:gelfitc}).} 
  \label{fig8}
\end{figure}

\section{Membranes near a gel transition}
\label{sec:gel}

Next, we consider the CD model with $\epsilon=1.3\,k_BT$, which in the symmetric case is located close to the liquid to gel transition of the membrane. The lipid diffusion coefficient for $\epsilon=1.3\,k_BT$ is almost an order of magnitude smaller than its corresponding value for the the fluid bilayer with $\epsilon=1.05\,k_BT$, discussed in section~\ref{sec:fluid}. We first consider the setting involving non-local rigidity, where we simulate two separate membrane segments that supposedly relax to a state without differential stress. The results from this set of simulations are summarized in fig.~\ref{fig7}. The elastic behavior displayed in this figure is quite similar to the one observed in figs.~\ref{fig5} and \ref{fig6} for the fluid membranes with the smaller value of $\epsilon=1.05\,k_BT$. The relaxed area, area elasticity modulus, and total bending rigidity [sub-figures~fig.~\ref{fig7}(a),(b),(c), respectively] show weak dependence on $\delta$, which should be largely related to non-quadratic corrections to the free energy Eq.~(\ref{eq:bilayelasticity3}), while the spontaneous curvature [sub-fig.~\ref{fig7}(d)] exhibits a clear linear increase with $\delta$.

To capture the behavior of a single bilayer, we turn to the set of MC simulations that allow transitions of lipids between the oppositely-curved segments, i.e., the setting without non-local rigidity. These simulations reveal a very distinct elastic behavior. Specifically, significant deviations from the description outlined by the elasticity theory in section~\ref{sec:theory} become noticeable for $\delta \geq 0.06$. This is demonstrated, for instance, in fig.~\ref{fig8}(a) showing the area free energy $F_1(A)$ obtained from the area distribution function $\Pi(A)$ via Eq.~(\ref{eq:fit1}). The black, red, and green symbols in fig.~\ref{fig8}(a) show $F_1(A)$ for $\delta=0.06,\ 0.08$, and  $0.10$, respectively. The computed free energies exhibit strong deviations from the quadratic form of Eq.~(\ref{eq:farea}). Two particularly interesting features emerge from the data: (i) The minimum of the free energy, corresponding to the relaxed area $A_0$, is reduced when the asymmetry parameter $\delta$ increases from 0.06 to 0.08, and (ii) there is a significant probability [associated with a relatively low free energy difference $F_1(A)-F_1(A_0)$] to find the membrane with a area $A<A_0$. These results suggest that the upper (compressed) monolayer may be undergoing a liquid to gel transition~\cite{vazques21,FOLEY20222997}. A strong indication for such a possible phase transition is the rapid decrease in the lipid diffusion coefficient measured in the upper layer, but not in the lower (dilated) layer, when $\delta$ is varied from 0.06 to 0.08 (i.e., concurrently with the decrease in $A_0$). An additional striking observation that points to the phase change comes from the curvature probability distribution, $\Psi(c)$, and the associated free energy, $F_2(c)$, which is plotted in fig.~\ref{fig8}(b). The results reveal that the the phase transition occurring at $\delta$ value between 0.06 and 0.08 involves not only a sudden change in the relaxed area $A_0$, but also a curvature instability. The bilayer state of mechanical equilibrium changes from flat to a state where the two segments have non-vanishing curvatures [which, by the construction of the configuration phase space in our simulations, are opposite to each other - see fig.~\ref{fig1}(b)]. From simple symmetry considerations it is clear that such a "buckled" state cannot be the equilibrium of a membrane with similar densities across each layer, and we readily conclude that the condensed layer does not gel uniformly, and the lipid densities at two oppositely-curved segments must be different.  

To write the bilayer elastic free energy, we (i) express the elastic free energies of the phase-separated segments, as the sum of quadratic area and curvature elasticities of their monolayers [compare with Eq.~(\ref{eq:bilayelasticity1})], then (ii) sum up the segments contributions: 
\begin{equation}    
    F=\sum_{\alpha=1,2}\sum_{i=1,2} 
    \left[\frac{K^{\alpha}_i(A^{\alpha}_i-A^{\alpha}_{i,0})^2}{2A^{\alpha}_{i,0}}+A^{\alpha}_i\frac{\kappa^{\alpha}_i(c^{\alpha}_i-C_{i,0}^{\alpha})^2}{2}\right],
    \label{eq:energymixed1}
\end{equation}
where the superscript $\alpha=1,2$ refers to, respectively, the segments with positive and negative curvatures, and the subscript $i=1,2$ corresponds to the upper and lower monolayers respectively. To express the free energy (\ref{eq:energymixed1}) in terms of bilayer area $A$ and the curvatures of the segments $\pm c$, we must use the following geometric relations: 
\begin{eqnarray}
    A^1_1=A^2_2=\frac{A(1+hc/2)}{2}, \label{eq:relapos} \\
    A^1_2=A_1^2=\frac{A(1-hc/2)}{2}, \label{eq:relaneg}
\end{eqnarray}
where $h$ is the distance between the monolayers [Eq.~(\ref{eq:avcurvature})]. Further assuming that $h\ll 1/c$, permits to replace $c^1_1\simeq c^1_2\simeq c$, and $c^2_1\simeq c^2_2\simeq -c$. To find the state of mechanical equilibrium of the tensionless bilayer, we must then minimize the free energy w.r.t.~$A$ and $c$ [which are now coupled through Eqs.~(\ref{eq:relapos}) and (\ref{eq:relaneg})], subject to the constraints that monolayer stresses are equal in both 
segments and opposite to each other, i.e., 
\begin{equation}
    K^1_1(A^1_1-A^1_{1,0})=K^2_1(A^2_1-A^2_{1,0})=K^1_2(A^1_{2,0}-A_2^1)=K_2^2(A^2_{2,0}-A_2^2).
    \label{difstressmixed}
\end{equation}
Also, we must take into account the fact that the lipids can move between the segments (despite of their low diffusivity), which imposes the following relationship between the relaxed areas: 
\begin{eqnarray}
    A^1_{1,0}/a^1_{1,0}+A^2_{1,0}/a^2_{1,0}=N_+, \label{eq:relaxedup} \\
    A^1_{2,0}/a^1_{2,0}+A^2_{2,0}/a^2_{2,0}=N_-, \label{eq:relaxeddown}
\end{eqnarray}
where $a^{\alpha}_{i,0}$ is the equilibrium area per lipid of the phase in the monolayer $i$ of segment $\alpha$.   

Due to the coupling between bilayer area $A$ and curvatures $\pm c$, the elastic free energy (\ref{eq:energymixed1}) is {\em not}\/ equal to the sum of the free energies displayed in fig.~\ref{fig8}: $F(A,c)\neq F_1(A)+F_2(c)$. Instead, we should regard $F_1(A)$ and $F_2(c)$ as operational free energies from which the effective bilayer moduli $K$ and $\kappa$ can be evaluated. Recent experiments found a considerable increase in the measured bending rigidity of substantially-asymmetric vesicles compared to their symmetric and weakly-asymmetric counterparts~\cite{C5CC00712G,C5CC10307J}. A sudden increase in $\kappa$ was also measured in MARTINI simulations of membranes with $N_+\neq N_-$~\cite{HOSSEIN2020624}. It was speculate that this observation may be related to the gel transition of the compressed layer and the reasonable expectation that the bilayer then becomes stiffer than in the liquid state. To look for this phenomenon in our simulation results, we must first discuss how $\kappa$ should be measured from the results in fig.~\ref{fig8}(b) for $\delta$ larger than the transition value. In this regime, the bending free energy $F_2(c)$ has two minima at $c=\pm C_0$, and $\kappa$ is measured by fitting $F_2(c)$ to
\begin{equation}
    F_2(c)=A_0\frac{\kappa}{2}(c-C_0)^2+E_2,
    \label{eq:gelfitc}
\end{equation}
close to one of the minima. Furthermore, one must measure $\kappa$ only along the stable $|c|>C_0$ branch of $F_2(c)$, and avoid the unstable branch $|c|<C_0$ representing flat transition states between the two equivalent sub-spaces of configurations obtained by reflection of the membrane w.r.t.~the vertical $z$-direction.  Such a fit to the results for $\delta=0.1$ is shown by the blue dashed curve in fig.~\ref{fig8}(b). Fig.~\ref{fig9} summarizes our measurements of $\kappa$ vs.~$\delta$. We find that $\kappa\sim 12.5\,k_BT$ for $\delta<0.06$, and $\kappa\sim 18\,k_BT$ for $\delta>0.08$. In the transition regime, the data does not fit well to Eq.~(\ref{eq:fcurve}) (for $\delta=0.06$), or Eq.~(\ref{eq:gelfitc}) (for $\delta=0.08$), and therefore $\kappa$ cannot be evaluated.    

\begin{figure}[t]
\centering
\includegraphics[height=5cm]{fig9.eps}
  \caption{The bending rigidity $\kappa$, and as a function of the asymmetry parameter $\delta$. The simulated bilayers have CD parameter $\epsilon=1.3\, k_BT$, and the measurements are conducted in the setting without non-local bending rigidity.} 
  \label{fig9}
\end{figure}

\begin{figure}[t]
\centering
\includegraphics[height=5cm]{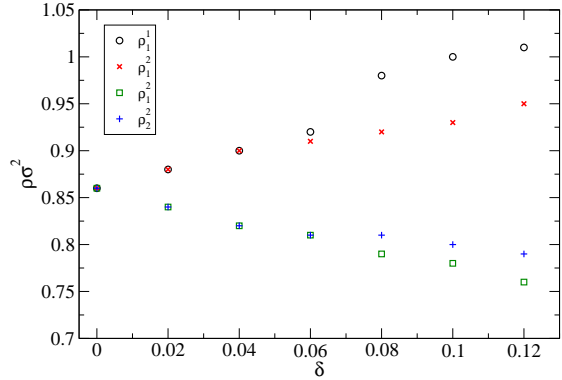}
  \caption{The average area density, $\rho^{\alpha}_i$, in the upper and lower layers (subscript $i=1,2$, respectively) of the segment with positive and negative curvature (superscript $\alpha=1,2$, respectively).} 
  \label{fig10}
\end{figure}

One must keep in mind that the results plotted in fig.~\ref{fig9} do {\it not}\/ show the bending rigidities of neither of the membrane segments, but the effective $\kappa$ of the whole membrane which is, roughly, the average of the segments rigidities. To take a closer look at the differences between the segments, we measured the average area densities, $\rho^{\alpha}_i$, in the different monolayers of the two segments. Similarly to Eq.~(\ref{eq:energymixed1}), we use the 
superscript $\alpha=1,2$ to denote the segments with positive and negative curvatures respectively, and the subscript $i=1,2$ to refer to the upper and lower monolayers respectively. The results are summarized in fig.~\ref{fig10}. For $\delta<0.06$, the densities in the condensed layer grow linearly with $\delta$. The densities decrease, by the same amount, in the dilated layer. These exactly opposite variations in the densities imply that the monolayers are oppositely strained, and that the strain is linear in $\delta$. Since the monolayers elastic free energy is quadratic the strain [see Eq.~(\ref{eq:bilayelasticity1})], the stress-strain relation of the monolayers in the small-$\delta$ regime is linear, which implies the monolayers are oppositely-stressed. 

For $\delta>0.06$, the data reveals a difference in the lipid densities at the upper monolayers of the positively ($\rho^1_1$) and negatively ($\rho^2_1$) curved segments. The latter continues with the linear dependence on $\delta$, characteristic of liquid membranes, while the former exhibits a discontinuity which is a signature of a gel transition. Quite remarkably, we observe that the phase separation occurring at the condensed layer, induces a density difference between the oppositely-curved segments of the dilated layer as well. This is surprising because the dilated layer is in a liquid state, and the uniformity of the stress in this layer suggests that the density must also be uniform, i.e., the same on both segments. A possible explanation for this seemingly odd observation is the coupling between monolayers, which has been completely ignored thus far, both in sections~\ref{sec:theory} and \ref{sec:gel}. Explicitly, lipids in each layer also interact with lipids in the opposite layer and, therefore, a phase separation in one layer may indeed lead to density variations in the other. Possibly-related effects originating in cross-layer interactions may be encountered in raft-forming biological membranes~\cite{WOS:000220358100004}, but this discussion in beyond the scope of this paper.          

\section{Discussion and conclusions}

The availability of simulation packages, like LAMMPS and GROMACS, enabling simple design and execution of Molecular Dynamics simulations, have contributed tremendously to the popularity of this computational method in many areas of molecular science and engineering. Regrettably, this has led to a decline in the usage of the alternative approach to molecular simulations, i.e., MC sampling, which has some unique capabilities. In this paper, we presented a new method for measuring the mechanical properties of an ultra CG bilayer membrane model with different lipid area densities in the two monolayers. The method exploits a useful feature of MC dynamics, which is to make statistical sampling of a partial configuration phase space. This is achieved by supplementing the Metropolis criterion for importance sampling with certain rejection rules. In the present study, these rules ensure that the initial asymmetry of the bilayer is preserved during the course of simulations. Partial statistical sampling provides us with the route to measure the area, $K$, and bending, $\kappa$, elastic moduli, as well as of the non-local bending modulus $\kappa_{\rm nl}$ that cannot be measured in full statistical sampling with PBCs.   

Starting from simulations of symmetric membranes with the same number of lipids in the upper ($N_+$) and lower ($N_-$) leaflets, we investigated how their mechanical behavior changes with the asymmetry parameter $\delta=(N_+-N_-)/(N_++N_-)$. We simulated membranes under two different conditions - one when the lipid area density in the symmetric case is much smaller than the density at the liquid to gel transition, $\rho_g$, and the other when it is only slightly smaller then $\rho_g$. In the former case, we find excellent agreement between the computational results and the elasticity theory for asymmetric liquid membranes, briefly reviewed in section~\ref{sec:theory}. The theory makes two simple assumptions: (i) The bilayer elastic free energy is the sum of the monolayers contributions. (ii) For each monolayer, the elastic free energy is written as the sum of area and bending quadratic elastic terms. From (i) and (ii) follows that the bilayer elastic free energy can be also written as the sum of quadratic terms in the strains associated with area density and curvature, with an additional non-local bending rigidity (or, equivalently, area-difference elasticity) term associated with the elastic frustration arising from the fact that it may not be possible to relax the areas of the two leaflets simultaneously. Non-local bending rigidity, however, has null to negligible impact on the mechanics of weakly fluctuating vesicles. This is the reason why $\kappa_{\rm nl}$ cannot be extracted from the spectrum of thermal fluctuations, and is only important (and measurable) in processes like fusion and budding involving significant changes in the geometry of the vesicle. One of the outcomes of the theory is the well known fact that the monolayers of a tensionless bilayer may not be tensionless themselves, but experience stresses that are similar in size and opposite in sign. Small deviations of the computational measurements from the predictions of the linear elasticity theory have been observed at high values of $\delta$, which can be attributed to corrections to the quadratic elastic free energies that are expected when the stresses in the monolayers grow. 

When the density of a symmetric membrane is only slightly smaller than $\rho_g$,  an increase in  the degree of asymmetry $\delta$, may cause the density of the compressed layer to exceed $\rho_g$. In contrast to symmetric membranes where both layers undergo a first order gel transition, in the substantially-asymmetric case only the compressed monolayer gels, and only in one of the two segments of our simulation membranes (i.e., at coexistence with a lower density in the other segment).  Furthermore, the coexistence of regions with different densities in the compressed layer may also induce correlated density variations in the dilated layer. The two bilayer segments with different monolayer densities, are mechanically coupled - the stresses in the upper layers of the segments must be identical, and (for a tenionless bilayer) be opposite to the stresses in their lower layers. The segments elasticities are characterized by different values of the elastic parameters, $a_0$ (area per lipid), $K$ (area elasticity modulus), $\kappa$ (bending modulus), and $C_0$ (spontaneous curvature).  The consequence of the mechanical coupling between the segments is a buckling instability into a new state of mechanical equilibrium, which is not flat as in the symmetric case. A theory for the elasticity of such inhomogeneous membranes is outlined in section~\ref{sec:gel}. Rather than attempting to minimize the elastic free energy of the asymmetric inhomogeneous membrane, we present an operational definition for the effective bending rigidity of the bilayer. Measurements of the effective $\kappa$ as a function of $\delta$, reveal that its value increases abruptly at the gelling-buckling phase transition. A rapid increase in the value of $\kappa$ has indeed been observed in experimental~\cite{C5CC00712G,C5CC10307J} and computational~\cite{HOSSEIN2020624} studies, and our study confirms that it is related to the stiffening of the membrane when it becomes partially gel-like.  

The bilayer membranes considered in this work are composed of a single type of lipid with different number of lipids in their leaflets. This type of asymmetry is far simpler than the compositional asymmetry of complex biological membranes that contain many different lipid species, and whose distribution within- and between-layers varies in time. Moreover, the bilayer in our simulations are also confined to have two segments of equal area with opposite curvatures. Yet, despite its simplicity, this model captures the mechanism of density-buckling instabilities originating from the mechanical coupling between the coexisting segments. We can now envision how this mechanism may be biologically important. Many cellular processes are initiated in liquid-ordered raft domains and, moreover, their progression involves variations in the compositions in the raft segments~\cite{SVIRIDOV2020598}. These density variations play a role not only in the biochemistry of cellular processes, but also in the biomechanics of correlated local curvature changes~\cite{C3SM51829A,curve-rafts18}. In the present model, the instability is driven by the liquid to gel first order transition and is, therefore, discontinuous in $\delta$. However, the change in the density and the coupled curvature variations may be continuous. The mechanism of density-curvature instability depends on the dissimilarity of the elastic parameters of different membrane regions. It is not limited to coexisting liquid and gel phases, but is also expected to affect the curvature of the liquid-ordered raft domains that are surrounded by the disordered liquid matrix of the cell~\cite{simons97}. Experiments with large unilamellar vesicles composed of mixtures of saturated and unsaturated lipids with cholesterol exhibit a rich phase diagram~\cite{FEIGENSON200947}, including also second-order~\cite{veach07} and possibly other phase transitions involving macro- or micro-phase separations~\cite{hirst11}, and even transitions from a macro- to micro-separation~\cite{GOH2013853,sarkar23}. Several theoretical frameworks involving similar mechanisms of curvature-density coupling have been proposed to explain the puzzle of micro-phase separation in lipid mixtures~\cite{schick12,schmid13}, which seems contradictory to the notion of thermodynamic phase separation. 
The molecular composition of biological membranes is far more complex, heterogeneous, and dynamic than of synthetic model mixtures. In such complex membranes, local density-curvature coupling can lead to many different curvature effects, some of which with relevance to specific biological processes. A major difference between large complex membranes and the simple CG model investigated  herein, is the fact that the later considers a bilayer with an identically vanishing average curvature with only two segments of opposite curvatures. The average curvature of the plasma membrane in not fixed, but its temporal variations are small because the plasma membrane includes multiple domains, whose sizes are much smaller then the cell size. Thus, on the cell-scale, the non-local curvature elasticity is unimportant, and its influence is limited only to the local and transient curvature variations. 

While this paper focuses on the connections between the lipidic heterogeneity and membrane asymmetry, it is important to remind that biological membranes are populated with embedded proteins and proteins that are associated with one of the leaflets. These proteins contribute to the membrane asymmetry in various ways. A prime example are ATP-dependent transporters, such as flippases, floppases, whose role in regulating lipid asymmetry has been already mentioned at the introduction of this manuscript~\cite{WOS:000276717600019,POMORSKI201669}. In addition to their direct impact on lipids, proteins also contribute to asymmetry through structural and functional adaptations~\cite{POGOZHEVA20132592}. The asymmetric shapes and orientations of transmembrane proteins, established during their directional insertion in the endoplasmic reticulum, not only create uneven functional domains across the bilayer, but also stabilize lipid asymmetry by interacting with specific lipid species~\cite{ilevental20,pabst2023}. Peripheral proteins further amplify these effects by selectively binding to lipids, influencing membrane curvature and enhancing structural integrity~\cite{PABST2024333}. Additionally, the attachment of the inner leaflet to the cytoskeleton and the outer leaflet’s interactions with extracellular matrix proteins or ligands significantly influence the asymmetric distribution of membrane components and the membrane spontaneous curvature~\cite{10.3389/fcell.2016.00155,lu2022}. These examples illustrate the complex interplay between lipids and proteins in achieving a stable yet adaptable asymmetric membrane architecture, underscoring the importance of also considering protein contributions when studying membrane asymmetry. We expect the growing research interest in asymmetric membranes to yield new exciting insights into the properties of such membranes, which might shed new light into the relationship between the asymmetric nature of biological membranes and their cellular functions.

\begin{acknowledgments}
This work was supported by the Israel Science Foundation (ISF), Grant No. 1258/22.
\end{acknowledgments}

\bibliography{aipsamp}

\end{document}